\documentstyle[preprint,aps]{revtex}
\begin{document}
\draft
\title{SPHALERONS AND ELECTROWEAK STRINGS} 

\author{Vikram Soni} \address{Theory group, National Physical Laboratory,\\
Dr.  K.  S.  Krishnan   Road,   New   Delhi   110012,   India.}  \maketitle

\begin{abstract} 
We show that the sphaleron  energy which  identifies  with the  (instanton)
potential  barrier for B-violation is reduced in the presence  (background)
of an  electroweak  Z-string.  We also  show that for  large  enough  Higgs
coupling,  $\lambda$,  the  sphaleron  energy  can go  negative.  For  such
$\lambda$,  electroweak  Z-strings can reduce their energy by  accumulating
sphaleron  bound  states.  This  further  endows them with  baryon  number.
Given our  approximation,  the value of $\lambda$ at which this occurs  is
rather large to be realistic for the standard electroweak model.  %

\end{abstract}

\section*{Introduction} 

There has been a lively  interest  of late in several  scenarios  of baryon
number (B) violation  with an eye to the  B-asymmetry  of the  universe.  A
particular   saddle-point   solution  of  the   electroweak   theory,   the
sphaleron\cite{ref1,ref2,ref3},  has  been  identified  with the  potential
barrier for the  instanton  process that takes one from one pure gauge (PG)
winding  number  vacuum to an adjacent  one.  Since the  instanton  process
changes the winding number by one unit and the anomaly translates this into
changes  of B, these  sphaleron  solutions  are  relevant  for B  violating
processes in the early universe.  %

More  recently,  vortex or string like  solutions  in the same  electroweak
model have been considered~\cite{ref4}.  Further , it has been shown that a
linked  configuration of loops of such strings carries winding number (also
termed Chern Simons  number, $N_{ cs}$)  \cite{ref5}.  Such  configurations
can then be broken and twisted into two  sphalerons  which can decay to the
vacuum of winding  number, $N_{cs} = 0$, providing a B violating  scenario.
This could have  relevance  to the B  asymmetry  of the  universe  as these
strings may have formed during the electroweak  phase transition  (EWPT) in
the evolution of the universe.  %

Much earlier Nambu~\cite{ref6} introduced these electroweak strings in  a
different context, while considering a dumbbell shaped object  in
the electroweak theory that has a monopole and an antimonopole at
the  extremities  joined by such a string or flux tube.  Such  an
object  could be stabilized by rotating it when  the  centrifugal
force  can balance the string tension which would  normally  contract the flux
tube to zero. Nambu's dumbbell is also one of  the
configurations encountered in the passage of twisted loop to  two
sphalerons~\cite{ref5}.

There  is a quite  different  development  that we  shift  to  now.  It was
pointed out by the  author~\cite{ref7}  that a sphaleron like configuration
which has an asymptotic magnetic field %
\begin{equation}
\left . B^{\text sp} \right| {_\infty} = \frac{4}{e}
\eta_{\text sp} \frac{\hat e_r \sin\theta}{r^2}
\end{equation}
can interact with an extended magnetic field so as to bring down the energy
of the sphaleron configuration very substantially This happens via the long
range attractive magnetic  interaction that is controlled by the parameter,
$B_0 R$ , where $B_0$ is the uniform  applied  external  magnetic field and
$R$ is the  typical  length  scale of the  region  over  which the field is
applied.  When  the  control  parameter  $B_0  R\sim   10^8$~gauss~cm,  the
attractive  magnetic  interaction can compete with the usual sphaleron mass
and  total  energy of the  sphaleron  configuration  comes  down to zero or
below.  This  observation  has import for  situations  which carry extended
magnetic fields, e.g.\ stars.  %

It seemed of interest to look upon the flux tubes of ~\cite{ref6} or the EW
strings of ~\cite{ref4}  and \cite{ref5} in this light.  However, there are
some    notable     differences.    The    strings    (flux    tubes)    of
{}~\cite{ref4}-\cite{ref6}  carry SU(2) flux.  Nambu considers strings with Z
flux as they are energetically favored and  ~\cite{ref4,ref5}  have strings
with Z ,W$^+$, W$_-$ flux, whereas  above  \cite{ref7}  we have  considered
real  magnetic  flux.  Thus, there cannot be any  long--range  interactions
with EW strings, since long range interactions are purely  electromagnetic.

The  range  of  interaction  for  the  Z  string  is  set  by  $1/  m_Z\sim
10^{-16}$cm.  We expect this is the typical  radius of the Z string.  Given
this  radius and the usual flux  quantum  carried by the  string we find an
average Z magnetic  field  $B^Z\sim  10^{24}$~gauss  for the Z string.  The
control parameter , $B^Z R\sim  10^8$~gauss~cm is of the same order as that
above , which  brought  sphaleron  energies down to zero.  Also, given that
the entire Z magnetic field acts inside of a radius  $1/m_Z$,  within which
the Z field is non-zero the lack of a long-range interaction may not scotch
all optimism.  %

We will address the  question of the  interaction  of a sphaleron  with a Z
string which may have been formed during the EWPT, to see if the  sphaleron
energy can be  markedly  lowered by its  interaction  with the string.  The
sphaleron energy is identified with the instanton  potential barrier and if
sufficiently  lowered could catalyse  B-violation.  We find, though not for
the expected  range of values of the higgs mass, that the sphaleron  energy
can even go negative  which implies that a sphaleron can bind to the string
and lower its energy.  %

\section*{The Z-string or the Nambu string}

The dumbbell object considered by  Nambu~\cite{ref6} is made up of an SU(2)
monopole  which is separated  from an SU(2)  antimonopole  with a flux tube
running along the axis.  In the limit that the separation  becomes large we
get a Z-string.  %

We  adopt  the  convention  in  ~\cite{ref6}.  The  relevant  part  of  the
electroweak     lagrangian     is
\begin{equation}     {\cal    L}    =
-\frac{1}{4}G_{\mu\nu}^a G^{\mu\nu a}-\frac{1}{4}F_{\mu\nu}^0  F^{\mu\nu 0}
-\frac{1}{2}   (D_\mu   \phi)^\dagger   D_\mu  \phi   -\frac{1}{8}\lambda^2
{(\phi^\dagger\phi -F^2)}^2
\end{equation}
where $m_W = gF/2$, $m_H=\lambda F$ and ${ F(VEV)} = 250$~Gev.  %

The Higgs configuration for the monopole is
\begin{equation}
\phi = F \left(
	\begin{array}{c}
		\cos (\theta /2)\\ \sin (\theta /2) e^{i\phi}
	\end {array}
	\right)
\end{equation}
where $\theta$  is the polar angle and $\phi$ the azimuthal angle with the
monopole at the origin.

The antimonopole is correspondingly
\begin{equation}
\overline\phi = F \left(
\begin{array}{c}
\sin (\theta/2)\\ \cos (\theta/2) e^{i\phi}
\end{array}
\right)
\end{equation}

Now, as  observed by Nambu  \cite{ref6},  in the  electroweak  theory,  the
monopole   has  a   spreading   SU(2)  flux  out  of  the  pole  given  by,
$-4\pi\eta/g$,  and a  semi-infinite  flux tube  attached to the pole which
carries a SU(2) flux, $-4\pi\xi /g$, out of the pole, with the constraint %
\begin{equation}
\eta + \xi  =  1
\end{equation}

     The  U(1)  field  is   sourceless   and   carries  a  spreading   flux
$4\pi\eta/g'$ and the same flux is returned via a semi-infinite  flux tube.

When we connect  the  monopole  with the  antimonopole  there is no flux at
infinity.  However, there is a flux tube  connecting  the two which carries
an SU(2) flux $-4\pi\xi/g$ and a U(1) flux  $-4\pi\eta/g'$.  We shall go by
the  electromagnetic  and Z field definitions given  in~\cite{ref6}  though
they are not unique.  %

The EM and Z flux carried by the flux tube are then given by
\begin{eqnarray}
\Phi_{\rm em} & = & -\sin\theta_W \Phi_{\rm SU(2)} + \cos\theta_W
\Phi_{\rm U(1)} =  \frac{4\pi}{e}(-\eta \cos^2\theta_W +
													\xi\sin^2\theta_{ W})\nonumber \\
\Phi_{\rm Z} & = & -\cos\theta_W \Phi_{\rm SU(2)} - \sin\theta_W
\Phi_{\rm U(1)}   =  \frac{4\pi}{e}(\sin\theta_W\cos\theta_ W)
\end{eqnarray}

where,  as we have defined earlier, \\ $\eta +\xi=1$, $e=g\sin\theta_W=g'
\cos \theta_W$ and $\tan \theta_W=g'/g$.

Notice,  that the Z flux,  $\Phi_Z$,  above  is  quantized  --- it does not
depend on $\eta$ or $\xi$,  whereas the EM flux,  $\Phi_{  em}$,  does.  As
Nambu  observes  a maximum  reduction  of flux  field  energy is  gained if
$\eta=\sin^2\theta_W$ or $\xi=\cos^2\theta_W$  leaving a purely Z flux tube
connecting the pole and the antipole.  This corresponds to a Z-string.  %

The energy per unit length of the flux tube is  calculated by assuming that
the Higgs  field is zero  inside the flux tube (of radius  $\rho$)  and the
fluxes above are averaged over the  cross-section  to give a uniform field.
\begin{equation}
\frac{E}{L}=\int_0^{2\pi}\int_0^\rho d\phi \rho' d\rho'
\left [ \frac{\lambda^2}{8} {(F^2)}^2 + \frac{1}{2} B^2_{\rm SU(2)} +
\frac{1}{2} B^2_{\rm U(1)}\right ]
\end{equation}
Now,
\begin{eqnarray}
\vec B_{\rm  SU(2)} & = & \Phi_{\rm SU(2)}/ \pi \rho^2 = - 4\pi\xi /g \pi
\rho^2 \cdot \hat e_Z \nonumber \\
\vec B_{\rm U(1)} & = & \Phi_{\rm  U(1)}/\pi \rho^2 = - 4\pi\eta /g' \pi
\rho^2 \cdot \hat e_Z \nonumber \\
\frac{E}{L} & = & \pi\rho^2
	\left [
		\frac{\lambda^2}{8} F^4 + 8
			\left \{
				\frac{\xi^2}{g^2} + \frac{\eta^2}{{g'}^2 }
			\right \}
		\frac{1}{\rho^4}
	\right ]
\end{eqnarray}
Minimizing with respect to  $\rho^2$
\begin{eqnarray}
\pi \left[  \frac{\lambda^2}{8} F^4 \right . & - & \left .
\frac{8}{\rho^4}\left\{ \frac{\xi^2}{g^2}
+ \frac{\eta^2}{g'^2} \right \} \right ]  = 0
\nonumber \\
\rho_{\lambda} & = & {\left( { \frac{8}{\lambda F^2}}\right )}^{1/2}
{\left [  \frac{\xi^2}{g^2} + \frac{\eta^2}{g'^2} \right ]}^{\frac{1}{4}}
\nonumber \\
\frac{E}{L} & = & 2\pi
\left [
	\lambda F^2
 	{
		\left \{
			\frac{\xi^2}{g^2} + \frac{\eta^2}{g'^2}
		\right \}
	}^{\frac{1}{2}}
\right ]
\end{eqnarray}
One further minimizes with respect to the parameter   $\xi$
This yields the same condition as above: $\xi=\cos^2\theta_W$ or
$\eta=\sin^2\theta_W$;
that is a purely  Z flux tube or string.
After minimization, the radius, $\rho_0(\lambda)$, and the energy per unit
length, $\epsilon(\lambda)$,  of the string are given by
\begin{eqnarray}
\rho_0(\lambda) & = & {\left ( \frac{8\cos\theta_W \sin\theta_W}{\lambda F^2
e}\right )}^{1/2} \nonumber \\
\epsilon(\lambda) & = & \frac{2\pi \lambda F^2 \cos\theta_W \sin\theta_W}{e}
\end{eqnarray}
The  Z  magnetic  field can be calculated from the  flux  on  the
assumption  that it is uniformly distributed over the  flux  tube
cross-section
\begin{equation}
\left| B_0^Z \right|
= \frac{\Phi_Z}{\pi\rho^2_0(\lambda)}
= \frac{4\pi}{e}\frac{\sin\theta_W \cos\theta_W}{\pi\rho_0^2(\lambda)}
=\frac{\lambda F^2}{2}
\end{equation}
For convenience we can express these quantities in terms of their
values  for   $\lambda = 1$ indicated by the subscript 1.
\begin{equation}
\begin{array}{cccccc}
\rho_0(\lambda) & = & \frac{1}{\sqrt\lambda} \rho_0(\lambda=1) & = &
\frac{1}{\sqrt\lambda}\rho_1 \\
\epsilon(\lambda) & = & \lambda \epsilon (\lambda=1) & = & \lambda\epsilon_1 \\
B_0^Z  & = & \lambda B_0^Z(\lambda=1) & = & \lambda\frac{F^2}{2}
\end{array}
\end{equation}

\section*{The  interaction of the sphaleron configuration with  the   Z-string}

The energy functional for the standard electroweak theory is
\begin{equation}
E=\int d^3 x \left[ \frac{1}{4} F_{ij}^0 F_{ij}^0 +
\frac{1}{4} G_{ij}^a G_{ij}^a + \frac{1}{2}(D_i \phi)^\dagger (D_i \phi) +
\frac{\lambda^2}{8} (\phi^\dagger\phi - F^2)^2 \right]
\end{equation}
We  shall  work with the sphaleron configuration  in  \cite{ref1,ref7}  The
asymptotic fields are given by
\begin{eqnarray}
\left . \phi\right|_ \infty & = &
					F \left( \begin{array}{c}
					\cos\theta \\
					\sin\theta e^{i\phi} \end{array}\right) \nonumber \\
\left . g {\cal A}_i^a \right|_ \infty & = &
\frac{2(1-\eta_{\text sp})}{r}\sin\theta (\hat e_\parallel)^a (\hat e_\phi)_i
-\frac{2}{r} \cos\theta (\hat e_\perp)^a (\hat e_\phi)_i
- \frac{2}{r} (\hat e_\phi ^a)(\hat e_\phi)_i \nonumber \\
\left . g'  {\cal A}_i^0\right|_{\infty}  & = &
\frac{2\eta_{\text sp}}{r}\sin\theta (\hat e_\phi)_i \nonumber \\
(\hat e_\parallel)&=& (\hat e_r)\cos\theta  +(\hat e_\theta)\sin\theta
\nonumber \\
(\hat e_\perp)&=& (\hat e_r)\sin\theta  -(\hat e_\theta)\cos\theta  \nonumber
\\
\end{eqnarray}

First, we point out,  that,  above, the  surviving  asymptotic  fields  are
electromagnetic  and proportional to $\eta_{\text  sp}$.  The U(1) field is
also explicitly  proportional to $\eta_{\text  sp}$.  The $\eta_{\text sp}$
dependent  asymptotic magnetic field of the sphaleron  configuration, which
was  crucial to provide  the long  range  attractive  interaction  with the
external  magnetic field is no more relevant for its  interaction  with the
Z-string as the latter does not carry the usual  magnetic  field.  In fact,
asymptotically, the $\eta_{\text sp}$ dependent part does not contribute to
the  interaction  of the  sphaleron  with the Z-string .  We may  therefore
restrict ourselves to the pure SU(2) sphaleron,  $\eta_{sp}=0$, to make our
estimates.  %

Note,  that  for  the usual magnetic field case  we  have  a
sphaleron  like configuration and not a solution as  observed  in
\cite{ref7}. In this case once $\eta_{ sp}=0$, we have the  exact
SU(2)  sphaleron  configuration,which  is   more  respectable. The
sphaleron asymptotic fields are,
\begin{eqnarray}
\left . \phi\right|_ \infty & = &
					F \left( \begin{array}{c}
					\cos\theta \\
					\sin\theta e^{i\phi} \end{array}\right) \nonumber \\
\left . g {\cal A}_i^a \right|_ \infty & = &
\frac{2}{r}\sin\theta (\hat e_\parallel)^a (\hat e_\phi)_i
-\frac{2}{r} \cos\theta (\hat e_\perp)^a (\hat e_\phi)_i
- \frac{2}{r} (\hat e_\phi ^a)(\hat e_\phi)_i \nonumber \\
\left . g'  {\cal A}_i^0\right|_{\infty}  & = & 0
\end{eqnarray}
We use the ansatz  \cite{ref1,ref2,ref3}  where the solution is obtained by
multiplying the asymptotic  fields, above, by appropriate  radial functions
which go to unity at $\infty$ and regulate the fields at the origin.  These
functions   are   $f(r)$  and  $h(r)$  for  the  higgs  and  gauge   fields
respectively.  %

Now, instead of the usual magnetic field background , we need to have the Z
magnetic  field  of  the  Z-string  as  the   background.  This  is  easily
accomplished.  The expressions for SU(2) and U(1) magnetic fields are %
\begin{eqnarray}
gB^a_i & =  &\frac{4 (\hat e_r)_i}{r^2}\left [ -h(r)(1-h(r)) \right ] \left [
\cos \theta {(\hat e_{\parallel})}^a + \sin\theta {(\hat e_{\perp})}^a \right ]
\nonumber \\
& &+\frac{2}{r} \frac{dh(r)}{dr} \left \{ \left [ {(\hat e_{\theta})}_i \left [
-\sin \theta {(\hat e_{\parallel})}^a + \sin\theta {(\hat e_{\perp})}^a \right
]
\right ] -
{(\hat e_{\phi})}_i {(\hat e_{\phi})}^a \right \} \nonumber \\
& & -e |B_0^Z| {(\hat e_{Z})}_i
{(\hat e_{\parallel})}^a \cot \theta_W  \Theta (\rho(\lambda ) -\rho )
\nonumber\\
g^ \prime B^0_i & = & -e |B_0^Z| {(\hat e_{Z})}_i \tan \theta_W  \Theta
(\rho(\lambda ) -\rho )
\end{eqnarray}

We shall use two ansatze developed by Manton and Klinkhammer~\cite{ref3}.

\noindent {\bf Ansatz (a)}\hfill

\begin{equation}
h^{(a)}(\xi) =
\left \{
	\begin{array}{cc}
		\left (\xi/\Xi \right )^2 &   \mbox{for $\xi <\Xi$} \\
				1                             & \mbox{for $\xi>\Xi$}
	\end{array}
\right .
\end{equation}
\begin{equation}
f^{(a)}(\xi) =
\left \{
	\begin{array}{cc}
		\xi/\Omega  &  \mbox{ for $\xi \leq\Omega$} \\
				1                             & \mbox{ for $\xi >\Omega$}
	\end{array}
\right .
\end{equation}
where $\xi$ is the  dimensionless  variable,  $gFr$, and $\Xi$ and $\Omega$
are the sizes of the  (dimensionless)  regions  outside  of which the gauge
field radial function,  $h(\xi)$, and the higgs radial function $f(\xi)$ go
to their asymptotic values respectively:  %

\noindent {\bf Ansatz(b)}\hfill
\begin{equation}
h^{(b)}(\xi) =
\left \{
	\begin{array}{cc}
		\xi^2/(\Xi^2 P')  &   \mbox{for $\xi \leq \Xi$} \\
				1- L(\xi)     &   \mbox{for $\xi>\Xi$}
	\end{array}
\right .
\end{equation}
where $P' = (1+4/\Xi)$ and
\begin{displaymath}
L(\xi) = \frac{4}{(4+\Xi)} \exp \frac{1}{2}(\Xi - \xi )
\end{displaymath}
and %
\begin{equation}
f^{(b)}(\xi) =
\left \{
	\begin{array}{cc}
		A\xi/\Omega  &  \mbox{for $\xi \leq\Omega$} \\
				1-[M(\xi)/\xi]   & \mbox{for $\xi >\Omega$}
	\end{array}
\right .
\end{equation}
where
\begin{equation}
A=\frac{\sigma\Omega+1}{\sigma\Omega+2}  \hspace{0.5in}{\rm and}\hspace{0.5in}
M(\xi)=
\frac{\Omega}{\sigma\Omega+2}\exp{\sigma(\Omega-\xi)}
\end{equation}
and in our convention $\sigma=\lambda/g$.

We now  turn  to the  energy  of the  sphaleron  in the  background  of the
Z-string (this excludes the energy of the flux tube).  This  corresponds to
a background Z magnetic field, $B_0^Z\hat e_Z$.  %

The energy  expression for the two ansatze that follows is given by the sum
of the  sphaleron  energy  in the  absence  of the  magnetic  flux, and the
interaction  energy, $E_{\rm int}$.  However, since, the interaction energy
is  proportional to the field,  $B_0^Z$, it is the modulus of $E_{\rm int}$
which is  relevant;  for its sign can always be made  negative  by choosing
$B_0^Z$  to be  appropriately  parallel  or  antiparallel  to the  positive
z-axis.  We then have for the energy $E$ %
\begin{equation}
E=E_0-\left | E_{\rm int} \right |
\label{eqnE}
\end{equation}

We shall give the  expression  for the first  ansatz (a) and  suppress  the
expression for the ansatz (b) for reasons of economy of space %
\begin{eqnarray}
E_0^{(a)} & = & \frac{4\pi F}{g}
	\left [
		\frac{26}{35}\frac{8}{\Xi} +
			\left \{
				\Xi
				\left [ \frac{8}{15} - \frac{\beta}{2} +
					\frac{4}{15}\beta^3 -
					\frac{2}{35} \beta^5
				\right ] \Theta(\Xi - \Omega)
			\right .
	\right . \nonumber \\
& + &
\left .
	\left . \Omega
		\left [
				\frac{1}{16} + \frac{16}{210}{\beta}^{-3}
		\right ]
		\Theta(\Omega-\Xi)
	\right \} +
	\frac{\lambda^2}{g^2}\frac{\Omega^3}{105}
\right ]
\end{eqnarray}
where $\beta=\Omega/\Xi$.

We shall, however, give the full expression for $E_{\rm int}$ for both the
ansatze. For

\noindent{\bf Ansatz(a)}\hfill
\begin{eqnarray}
E_{\rm int}^{(a)}  & = &  - 4\pi\Gamma\Xi
\left [
	\left \{
		\frac{16}{45} - \left ( 1 + \frac{1}{3}\alpha^2\right )\gamma +
		\gamma^3  \left ( \frac{4}{9}+ \frac{\alpha^2}{3}
		\right ) + \frac{1}{5}\gamma^5 +
		\frac{4}{3}\alpha^3\arctan \left (\frac{\gamma}{\alpha}\right )
	\right \}
	 \Theta (\Xi-r_0)
	\right . \nonumber \\
	&  & + \left . \frac{16}{45}  \Theta (r_0 - \Xi)
\right ]
\end{eqnarray}
where $r_0 = g v \rho_0$, $\alpha = r_0/\Xi$, $\gamma=\sqrt{1-\alpha^2}$ and
\begin{equation}
\Gamma =  \frac{Fe}{g^3} \lambda \cot \theta_W
\end{equation}

\noindent{\bf Ansatz(b)}\hfill
\begin{equation}
E^{(b)}_{\rm  int} = E_1 \Theta (\Xi - r_0) + E_2 \Theta (r_0 - \Xi)
\end{equation}
where
\begin{eqnarray}
E_1 & = & -4\pi\Gamma
\left \{
	\left (
		\int_\Xi^\infty dz \int_0^{r_0^2} d(\rho^2) +
		\int_{\sqrt{\Xi^2 - r_0^2}}^\Xi dz
		\int_{\Xi^2 - z^2}^{r_0^2} d(\rho^2)
	\right )
	\right . \nonumber \\
	& &	\left [
			\frac{L(1-L)(-z^2)}{(z^2+\rho^2)^2}
			+ \frac{1}{4} \frac{L}{\sqrt{z^2 +\rho^2}}\frac{\rho^2}{(\rho^2 + z^2)}
		\right ]
\nonumber  \\
&   & +
	\frac{\Xi}{P'}
	\left [
		\frac{5}{9} + \frac{1}{3}\left (\frac{1}{P'} -1\right ) -
		\frac{1}{5P'} - \left (1 + \frac{\alpha^2}{3}\right )\gamma
		\right .
		\nonumber  \\
& &	\left . \left .	+	\gamma^3
	\left \{
		\frac{4}{9} - \frac{1}{3}
						\left (\frac{1}{P'} -1
						\right ) +
		\frac{\alpha^2}{3P'}
	\right \} +
		\frac{1}{P'}\frac{1}{5} \gamma^5
		+\frac{4}{3}\alpha^3 \arctan
						\left (\frac{\gamma}{\alpha}
						\right )
	\right ]
\right \}
\end{eqnarray}
and
\begin{eqnarray}
E_2  & = & -4\pi\Gamma
\left \{
	\left (
		\int_\Xi^\infty dz \int_0^{r_0^2} d(\rho^2) +
		\int_0^\Xi dz \int_{\Xi^2 - z^2}^{r_0^2} d(\rho^2)
		\right . \right . \nonumber \\
& &\left . \left .
		\left [
			\frac{L(1-L)(-z^2)}{{(z^2+\rho^2)}^2}
			+  \frac{1}{4} \frac{L}{\sqrt{z^2 +\rho^2}}\frac{\rho^2}{(\rho^2 + z^2)}
		\right ]
		+  \frac{\Xi}{P'}
		\left [
			\frac{5}{9} + \frac{1}{3}\left (\frac{1}{P'} -1\right ) - \frac{1}{5P'}
		\right ]
	\right )
\right \}
\end{eqnarray}

where   $r_0   =   3.35g/   \lambda^{1/2}$,   using   the   expression   for
$\rho_0(\lambda)$  from  the  previous  section.  We use  $g^2 =  0.4$  and
$\cot^2\theta_W= 1.83$.  $F=250$GeV.  %

Finally, on minimizing $E$ with respect to $d$ and substituting  this value
of $d$ in the energy we get the energy of the  sphaleron in the  background
of the Z flux tube,  $E_{\text  sp}$.  These are  displayed  in the tables.
For   comparison  we  display  the  usual   sphaleron   energies,   $E_{\rm
sp}^0$~\cite{ref3}, in the absence of the flux tube, as well.  %

For both the ansatze, the  interaction  energy,  $E_{\rm int}$, in $E_{sp}$
changes  sign as we go from large  $\lambda > 2$ to  $\lambda < 2$.  In our
convention  the sign of $E_{\rm  int}$ for small  $\lambda$ is positive and
that for large  $\lambda$  negative.  Both  ansatze  show the presence of a
minimum  in  $E_{sp}$  (expected)  around  $\lambda  \sim 0.1-  0.2$  where
$E_{sp}$  is just  slightly  lower  than in the  absence of  interaction  (
$\lambda = 0$ ).  As  $\lambda$  increases we move to a maximum in $E_{sp}$
around  $\lambda\sim  2$ , and then for larger  $\lambda$,  $E_{sp}$ starts
going  down.  For  ansatz  (a)  (see  table 1)  $E_{sp}$  is  monotonically
decreasing  as  $\lambda$  increases  and  goes  asymptotically  to  $\sim 2$
as
$\lambda\rightarrow\infty$.  For ansatz (b) (see table 2)  $E_{sp}$  falls much
 faster with
increasing  $\lambda$  going  to zero  for  $\lambda\sim  75$ and  becoming
negative for higher  $\lambda$.  As we have already  indicated  the sign of
$E_{int}$  changes  around  $\lambda$  between  2 and 5 but  since  only $|
E_{int} |$ occurs in the expression for  $E$ (Eq.~\ref{eqnE}), this sign is
not significant.  %

  Let us take up some of the shortcomings of this analysis which end up
overestimating the sphaleron energy in the string background.
  The sphaleron has a size in terms of the Higgs field. This is the radial
distance from the origin at which the Higgs field reaches its normal VEV.The
sphaleron energy (Eq.13) is composed of three positive definite terms ( leaving
out the interaction term )
 (a) the gauge field energy ,(b) the gradient energy for the Higgs field, and
(c) the Higgs potential term.
  The term (c) is a measure of the the energy due to the HIggs departing from
its normal VEV. Recall that the string was constructed by minimizing the sum of
the energy of emptying out the Higgs field from the flux tube and spreading out
the Z field uniformly in the crosssection of the flux tube.

If the sphaleron is contained in the flux tube then we have already accounted
for the energy of emptying out the Higgs field in this region. There is no need
for us to include the terms (b) and (c) above. Correcting this will reduce the
energy of the sphaleron substantially.

If on the other hand the sphaleron size is larger than the flux tube radial
size, then we need to add (b) and (c) above only in in the region exterior to
the flux tube.So we have overestimated the sphaleron energy as we included (b)
and (c) inside the the fluxtube as well.

  First, we provide a very simple illustration of the effect considered in this
paper. Consider the configuration with $\Omega$ = $\Xi$ = $r_0$ , in the
parameterization of Ansatz(a).In view of the foregoing this means that the
contributions of the terms (b) and (c) may be dropped in (Eq.23),as the
sphaleron is inside the flux tube. The corrected sphaleron energy is
\begin{equation}
     E_0^{(a)} =  \frac{4\pi F}{g}\frac{26}{35}\frac{8}{r_0}
\end{equation}

The interaction energy (Eq.24) is
\begin{equation}
    E_{ int}^{(a)} =  -4\pi\Gamma r_0\frac{16}{45}
\end{equation}
and we find the ratio    $  \frac{E{int}^{(a)}}{E_0^{(a)}}
\sim \frac{-3}{8} $ .

 This shows that the sphaleron energy is already reduced by a factor  5/8
through its attractive interaction with the EW string, even for this simplified
configuration which does not minimize the total energy, $E$.

Now, on including this correction we find that the value of $\lambda$ at which
the $ E_{sp}\leq 0$, ( Ansatz(b)), comes down to $\lambda\sim 60$ as compared
to$\lambda\sim 80$ , in the absence of the correction.This is already a
substantial lowering.
    We do not exhibit this correction for the whole table but just quote the
new value of $\lambda \sim 60 $ at which $  E_{sp}\leq 0  $.

We have used the  parametrization~\cite{ref3}  to calculate  the  sphaleron
energy.  Also, the energy, radius and Z magnetic field for the Z-string are
estimated  as in  Ref.~\cite{ref6}  without  using the exact  solutions  in
\cite{ref4} and  \cite{ref5}.  To get reliable numbers an exact solution is
required.  In  this  case  we  expect   that   $E_{sp}$   will   come  down
substantially and go to zero for smaller $\lambda$ than those quoted above.

\section*{Conclusions}
								We have  found  that  in the  background  of the EW
Z-string, $E_{sp}$, that is , the instanton potential barrier relevant to B
violation is reduced by  interaction.  In terms of catalysing  B-violation,
however, the reduction of the barrier becomes  significant  only for rather
large  $\lambda\sim  60$ , when  the  barrier  goes  to 0.  This  value  of
$\lambda$  is  unrealistic  for the  standard  EW  model  where  we  expect
$\lambda\sim 1$.  %

   When $E_{sp}< 0$ we have a  qualitatively  new situation.  The sphaleron
no longer has the interpretation of an instanton potential barrier.  It now
has an energy that is lower than the winding number (PG) vacua whose energy
is not  affected  by the flux tube.  Such a sphaleron  can bind to the flux
tube/string  and  lower  its  energy;  one  thus  expects  that  it will be
spontaneously produced on the flux tube.  This happens for $\lambda \geq 60
$.  In this regime,  $E_{sp} < 0$ , the ground state string  solution  will
have sphaleron bound states beading on it.This is a new string solution.  %

   Furthermore,  sphaleron  bound states would endow the EW Z-strings  with
   winding  number ( $N_{cs}$ ) and  consequently  baryon  number.  If such
   strings were created during the EWPT they would decay to the vacuum well
   after the EWPT  giving  rise to B  violation  and a B  asymmetry  in the
   presence  of  CP  violation.  This  could  yield  some  of  the  effects
   considered in \cite{ref5} without any looping and twisting.  %

   It is  intriguing  that if we had a flux  quantum an order of  magnitude
   larger  (than it  actually  is) for the EW string  we would  have  found
   $E_{sp}= 0$ for  $\lambda\sim 1$!  Therefore it is important to consider
   an exact string  solution where the magnetic field is much higher at the
   origin and falling  exponentially  out in contrast to the  averaged  out
   field we have used in Sec.2.  Also, an exact numerical  solution for the
   sphaleron  instead of the  ansatze we have used is  important.It  is not
   clear how sensitive our results are to these  details.It is evident that
   such a program  must be  importantly  and  urgently  carried  out to get
   definitive answers.  %

\begin{table}
\begin{tabular}{|c|c|c|c|c|c|c|c|}
$\lambda$ & $E_{sp}^0$ & $\Xi^0$  & $\Omega^0$ &  $E_{sp}$ & E$_{\rm int}$ &
$\Xi$  & $\Omega$ \\ \hline
0.001 & 2.40 & 4.95 & 4.8 & 2.399 & 0.002 & 4.95 & 4.8 \\
0.1 & 2.43 & 4.81 & 4.53 & 2.18 & 0.262 & 5.31 & 4.98 \\
0.2 & 2.48 & 4.56 & 4.05 & 2.015 & 0.469 & 4.77 & 4.209 \\
1.0 & 2.92 & 3.61 & 1.99 & 2.26 & 0.92 & 2.48 & 1.66 \\
2.0 & 3.17 & 3.41 & 1.19 & 2.62 & 1.105 & 1.92 & 1.02 \\
5.0 & 3.38 & 3.34 & 0.51 & 3.08 & -- 0.33 & 3.69 & 0.52 \\
10.0 & 3.47 & 3.34 & 0.263 & 2.94 & -- 0.535 & 3.4 & 0.26 \\
100.0 &  3.55 & 3.34 & 0.0264 & 2.46 & -- 1.187 & 2.85 & 0.026 \\
10000.0 & 3.56 & 3.33 & 0.00053 & 2.03 & -- 2.035 & 1.97 & 0.00026 \\
\end{tabular}
\caption{}
\end{table}

\begin{table}
\begin{tabular}{|c|c|c|c|c|c|c|c|}
$\lambda$ & $E_{sp}^0$ & $\Xi^0$  & $\Omega^0$ &  $E_{sp}$ & $E_{int}$ & $\Xi$
& $\Omega$ \\ \hline
0 & 1.566 & 2.66 & 2.60 & 1.566 & 0 & 2.66 & 2.60 \\
0.1 & 1.67   & 2.15  & 2.37  & 1.508 & 0.154 & 2.5 & 2.59 \\
0.2  & 1.77 & 1.81 & 2.05 & 1.54 & 0.47 & 2.25 & 2.30 \\
0.5  & 1.95 & 1.39 & 1.60 & 1.709 & 0.244 & 1.5 & 1.65 \\
1.0  & 2.13 & 1.10 & 1.19 & 1.928 & 0.174 & 1.2 & 1.22 \\
2.0  & 2.32 & 0.892 & 0.795 & 2.309 & 0.009 & 0.9 & 0.8 \\
10.0  & 2.61 & 0.733 & 0.21& 1.827 & -- 0.945 & 0.225 & 0.198 \\
80.0 & 2.708 & 0.729& 0.026 & -- 0.06 & -- 3.343 & 0.007 & 0.026 \\
100.0 & 2.711 &0.729 & 0.021 & -- 0.324 & --3.634 & 0.0042 & 0.021\\
\end{tabular}
\caption{}
\end{table}
\newpage
\centerline {Table Caption(s)}

\begin{itemize}
\item[Table I.]
 Ansatz(a):  Parameters   for  the   sphaleron   interaction   with  the  Z
                 string/flux  tube.  The  superscript  $^0$  indicates  the
                 values  of the  variables  for the case of no  interaction
                 which  match with those of  Ref.~\protect\cite{ref3}.  The
                 units of energy are ($4\pi F \ v/g$) again  conforming  to
                 the  convention  used  in   Ref.~\protect\cite{ref3}.  The
                 interaction  energy changes sign around  $\lambda\sim  2$,
                 though it is only the  modulus of \protect  $E_{\rm  int}$
                 which is relevant for \protect $E_{\rm sp}$ %
\item[Table II.]
Ansatz(b): Table caption same as for Table I.
\end{itemize}

\begin{references}

\bibitem{ref1}{V. Soni, Phys.\ Lett.\ {\bf B93}(1980)101.}
\bibitem{ref2}{N.S. Manton, Phys.\ Rev.\ {\bf D28} (1983)2019.}
\bibitem{ref3}{F. R. Klinkhammer and N.S. Manton, Phys.\ Rev.\ {\bf
D30}(1984)2212.}
\bibitem{ref4}{T. Vachaspati, Phys.\ Rev.\ Lett.\ {bf 68}(1992)1977,
T. Vachaspati, Tufts University preprint TUTP-92-12(1992)}
\bibitem{ref5}{T. Vachaspati, ``Electroweak String configurations with  Baryon
Number'', Tufts University preprint(1992)}
\bibitem{ref6}{Y. Nambu, Nucl. Phy. {\bf B130} (1977)505.}
\bibitem{ref7}{V Soni, ``A Baryon number violating classical  configuration
in the presence of external magnetic fields'', National  Physical
Laboratory (NPL) preprint (1994).}
\bibitem{ref8}{V.  Soni and S. Lamba, same title, NPL preprint, NPL-TH(P)-2
1991.}
\end{references}
\end{document}